\documentclass[aps,pra,twocolumn,superscriptaddress,showpacs]{revtex4}
\usepackage{mathrsfs}
\usepackage{graphicx}
\usepackage{amsfonts}
\usepackage{amsmath}
\usepackage{amssymb}

\begin{document}

\title{Quantum Nonlocality Enhanced by Homogenization}

\author{Xu Chen}
 \affiliation{Theoretical Physics Division, Chern Institute of Mathematics, Nankai University,
 Tianjin 300071, People's Republic of China}

\author{Hong-Yi~Su}
\email{hysu@mail.nankai.edu.cn}
 \affiliation{Theoretical Physics Division, Chern Institute of Mathematics, Nankai University,
 Tianjin 300071, People's Republic of China}

\author{Zhen-Peng~Xu}
 \affiliation{Theoretical Physics Division, Chern Institute of Mathematics, Nankai University,
 Tianjin 300071, People's Republic of China}

\author{Yu-Chun Wu}
 \affiliation{Key Laboratory of Quantum Information, University of Science and Technology of China, 230026 Hefei, People's Republic of China}

 \author{Jing-Ling~Chen}
 \email{chenjl@nankai.edu.cn}
 \affiliation{Theoretical Physics Division, Chern Institute of Mathematics, Nankai University,
 Tianjin 300071, People's Republic of China}
 \affiliation{Centre for Quantum Technologies, National University of Singapore,
 3 Science Drive 2, Singapore 117543}

\date{\today}

\begin{abstract}
Homogenization proposed in [Y.-C Wu and M. \.Zukowski, Phys. Rev. A 85, 022119 (2012)] is a
procedure to transform a tight Bell inequality with partial correlations into a full-correlation form
that is also tight. In this paper, we check the homogenizations of two families of $n$-partite  Bell inequalities: the
Hardy inequality and the tight Bell inequality without quantum violation. For Hardy's inequalities, their homogenizations bear stronger quantum
violation for the maximally entangled state; the tight Bell inequalities without quantum violation give the boundary of quantum and supra-quantum, but their homogenizations do not have the similar properties. We find  their homogenization are violated by  the maximally entangled state.  Numerically computation shows the the domains of quantum violation of homogenized Hardy's inequalities for the generalized GHZ states are smaller than those of Hardy's inequalities.
\end{abstract}

\pacs{03.65.Ud, 03.30.+p, 03.67.-a}

\maketitle

\section{Introduction}
The problem of the possibility of a local realistic interpretation of quantum mechanics was first addressed in the discussion between Einstein, Podolsky, Rosen (EPR)~\cite{EPR} and Bohr~\cite{Bohr}. In order to settle down the philosophical  debate, Bell proposed an experimental scheme in 1964~\cite{Bell}. Bell's seminal paper contains an inequality, which holds for local realistic correlations but can be violated by quantum mechanical correlations. In the work of Clauser, Horne, Shimony, and Holt (CHSH)~\cite{CHSH} a new inequality was derived, which, comparing with the original Bell's expression, can be more applicable to real experimental setups. An inequality with  lower-order (partial) correlations is usually called a 
Clauser-Horne-type (CH-type) Bell inequality, whereas an inequality involving only highest-order (full) correlations is usually referred to as a CHSH-type Bell inequalities.

To identify quantum nonlocality (QN), one needs a set of complementary observables for each party. It is anticipated that the ability to detect QN becomes stronger as the number of observables (settings) increases. For instance, the 2-setting CHSH inequality determines the visibility for the Werner state as $1/\sqrt{2}$, while a 465-setting inequality~\cite{CHSH2} decreases the visibility to 0.7056, which is slightly smaller than $1/\sqrt{2}$.

On the other hand, although the CHSH inequality can detect the QN for all two-qubit pure entangled states, for more parties it is not desirable to invoke the full-correlation inequality, like the CHSH inequality. In fact, the first Bell inequality to identify QN for the whole domain of the generalized Greenberger-Horne-Zeilinger (GHZ) state $\cos\theta|000\rangle+\sin\theta|111\rangle$ is comprised of partial correlations.

In general, in constructing Bell inequalities there is a trade-off between full correlations and the ability to detect QN for more generalized states. Wu and \.{Z}ukowski~\cite{Yu-Chun Wu1} show how to transform a CH-type inequality into a CHSH-type inequality, the tightness being preserved. In this paper, we use the procedure of homogenization to compare the quantum violation for various states. Through the homogenization, the setting for each party is increased by one. This strengthens the quantum violation for the maximally entangled state, while the violation regimes for the nonmaximally entangled state could be narrowed.

The organization of this paper is as follows. In Sec. II, we briefly review the procedure of homogenization. Then, in Sec. III, we focus on the generalized GHZ state and compare its quantum violation of the Hardy inequality before and after the homogenization. Likewise, in Sec. IV we study the influence of the homogenization on a family of tight Bell inequalities without quantum violation. We discuss the results in Sec. V and propose a possible reason for the enhanced QN by the homogenization.

\section{BRIEF REVIEW ON HOMOGENIZATION}
In general, the CHSH- and CH-type inequalities read
\begin{eqnarray}
\langle\sum_{ij}^{}{\omega }_{ij}{a}_{i}{b}_{j}\rangle\leq1,\label{CHSH}
\end{eqnarray}
\begin{eqnarray}
0\leq I(\hat{a},\hat{b})=\langle c+\sum_{i}^{}{\alpha }_{i}{a}_{i}+\sum_{j}^{}{\beta }_{j}{b}_{j}+\sum_{ij}^{}{\gamma }_{ij}{a}_{i}{b}_{j}\rangle\;\;\;\nonumber\\
\leq M,\label{CH}
\end{eqnarray}
where ${\omega }_{ij}$, $\alpha_{i}$, $\beta_j$ and $\gamma_{ij}$ are real coefficients, $c$ is a real constant, $M$ is the classical upperbound of (\ref{CH}), and $a_i,b_j$ are dichotomic observables taking values $\pm1$. Wu and \.{Z}ukowski~\cite{Yu-Chun Wu1} showed that (\ref{CH}) can be transformed into (\ref{CHSH}) by homogenization, and that if (\ref{CH}) is tight then (\ref{CHSH}) is also tight. Specifically, the homogeneous expression is obtained as
\begin{eqnarray}
H(I)=\langle c'{a}_{0}{b}_{0}+\sum_{i}^{}{\alpha }_{i}{a}_{i}{b}_{0}+\sum_{j}^{}{\beta }_{j}{b}_{j}{a}_{0}+\sum_{ij}^{}{\gamma }_{ij}{a}_{i}{b}_{j}\rangle,
\end{eqnarray}
with
\begin{eqnarray}
-\frac{M}{2}\leq \frac{1}{{a}_{0}{b}_{0}}H(I)\leq \frac{M}{2},\;\;\;\;c'=c-M/2.
\end{eqnarray}
\par

\section{Hardy's nonlocality inequalities}
Cereceda~\cite{Hardy} extend Hardy's nonlocality proof for two spin-1/2 particles~\cite{Hardy2} to the case of $n$ spin-1/2 particles configured in the generalized GHZ state. We now show that the maximal quantum violation of the homogenized Hardy's inequality is stronger than Hardy's original inequality. The $n$-qubit CH-type Hardy's inequality reads~\cite{Hardy}
%\begin{equation}
%\begin{split}
%P({\mathcal{\mathcal{U}}}_{1}{\mathcal{U}}_{2}{\mathcal{U}}_{3}\ldots{\mathcal{U}}_{n}|+++\ldots+)\\
%\leq P({\mathcal{D}}_{1}{\mathcal{D}}_{2}{\mathcal{D}}_{3}\ldots{\mathcal{D}}_{n}|---\ldots-)\\
%+P({\mathcal{D}}_{1}{\mathcal{U}}_{2}{\mathcal{U}}_{3}\ldots{\mathcal{U}}_{n}|+++\ldots+)\\
%+P({\mathcal{U}}_{1}{\mathcal{D}}_{2}{\mathcal{U}}_{3}\ldots{\mathcal{U}}_{n}|+++\ldots+)\\
%+\ldots+P({\mathcal{U}}_{1}{\mathcal{U}}_{2}{\mathcal{U}}_{3}\ldots{\mathcal{D}}_{n}|+++\ldots+)
% \end{split}
%\end{equation}
\begin{eqnarray}
p(0_1 0_2 0_3\cdots 0_n|0_1 0_2 0_3\cdots 0_n)\nonumber\\
\leq p(1_1 1_2 1_3\cdots 1_n|1_1 1_2 1_3\cdots 1_n)\nonumber\\
+p(0_1 0_2 0_3\cdots 0_n|1_1 0_2 0_3\cdots 0_n)\nonumber\\
+p(0_1 0_2 0_3\cdots 0_n|0_1 1_2 0_3\cdots 0_n)\nonumber\\
+\cdots+p(0_1 0_2 0_3\cdots 0_n|1_1 0_2 0_3\cdots 1_n),\label{hardy-n}
\end{eqnarray}
where $p(i_1 i_2 i_3\cdots i_n|j_1 j_2 j_3\cdots j_n)$ denotes the joint probability of obtaining result $i_k$ under setting $j_k$ for the $k$-th qubit, with $i_k,j_k=0,1$ and $k$ running from $1$ to $n$. In the following context the subscript for each party could be omitted with no confusion.
%$\mathcal{U}_i$ and $\mathcal{D}_i$ denote two distinct settings of the $i$-th qubit, $\pm$ denote two measuring results, and $P({\mathcal{\mathcal{U}}}_{1}{\mathcal{U}}_{2}{\mathcal{U}}_{3}\ldots{\mathcal{U}}_{n}|+++\ldots+)$ is the joint probability of obtaining result $+$ under setting $U$ of the first qubit, and $+$ for the second, etc., and similarly for the others.
Let us rewrite (\ref{hardy-n}) in the form of correlations. For $n=3$,
\begin{equation}
\begin{split}
|5-{A}_{1}-{B}_{1}-{C}_{1}-{A}_{2}{B}_{1}-{A}_{2}{C}_{1}-{A}_{1}{B}_{2}-{B}_{2}{C}_{1}\\
-{A}_{1}{C}_{2}-{B}_{1}{C}_{2}-{A}_{2}{B}_{2}-{A}_{2}{C}_{2}-{B}_{2}{C}_{2}+{A}_{1}{B}_{1}{C}_{1}\\
+{A}_{2}{B}_{2}{C}_{2}-{A}_{2}{B}_{1}{C}_{1}-{A}_{1}{B}_{2}{C}_{1}-{A}_{1}{B}_{1}{C}_{2}|/8\leq 1,\label{Hardy3}
 \end{split}
\end{equation}
where $A_1=p(0_1|0_1)-p(1_1|0_1)$ and $A_2=p(0_1|1_1)-p(1_1|1_1)$ are observables for the first party, and likewise for $B_i,C_j$.
It is violated by the three-qubit GHZ state by a factor of 1.1755, and it is violated maximally by the state $0.8393|000\rangle+0.5435|100\rangle+0.01283|101\rangle-0.0009074|110\rangle+0.002652|111\rangle$ by a factor of 1.236. In presence of noise, the state is a mixed state defined by $\rho=V|GHZ\rangle \langle GHZ|+(1-V)\rho_{\rm{noise}}$, where $V$ is called visibility, $\rho_{\rm{noise}}=\frac{1}{8}\mathbb{I}^{\otimes 3}$ for three qubits, and $\mathbb{I}$ is the $2\times2$ identity matrix. For the GHZ state, the threshold visibility $V$ is 0.6812.
The homogenized inequality can be written as
\begin{equation}
\begin{split}
|5{A}_{0}{B}_{0}{C}_{0}-{A}_{1}{B}_{0}{C}_{0}-{A}_{0}{B}_{1}{C}_{0}-{A}_{0}{B}_{0}{C}_{1}-{A}_{2}{B}_{1}{C}_{0}\\
-{A}_{2}{B}_{0}{C}_{1}-{A}_{1}{B}_{2}{C}_{0}-{A}_{0}{B}_{2}{C}_{1}-{A}_{1}{B}_{0}{C}_{2}-{A}_{0}{B}_{1}{C}_{2}\\
-{A}_{2}{B}_{2}{C}_{0}-{A}_{2}{B}_{0}{C}_{2}-{A}_{0}{B}_{2}{C}_{2}+{A}_{1}{B}_{1}{C}_{1}+{A}_{2}{B}_{2}{C}_{2}\\
-{A}_{2}{B}_{1}{C}_{1}-{A}_{1}{B}_{2}{C}_{1}-{A}_{1}{B}_{1}{C}_{2}|/8\leq 1,\label{G-Hardy3}
 \end{split}
\end{equation}
It is violated maximally by the three-qubit GHZ state by a factor of 1.8 (see Fig.~1), the threshold visibility $V$ is 0.5556.

For $n=4$,
\begin{equation}
\begin{split}
|20-2({A}_{1}+{B}_{1}+{C}_{1}+{D}_{1})-{A}_{1}{B}_{1}-{A}_{1}{C}_{1}-{A}_{1}{D}_{1}\\
-{B}_{1}{C}_{1}-{B}_{1}{D}_{1}-{C}_{1}{D}_{1}-{A}_{2}{B}_{1}-{A}_{2}{C}_{1}\\
-{A}_{2}{D}_{1}-{A}_{1}{B}_{2}-{B}_{2}{C}_{1}-{B}_{2}{D}_{1}-{A}_{1}{C}_{2}\\
-{B}_{1}{C}_{2}-{C}_{2}{D}_{1}-{A}_{1}{D}_{2}-{B}_{1}{D}_{2}-{C}_{1}{D}_{2}\\
-{A}_{2}{B}_{2}-{A}_{2}{C}_{2}-{A}_{2}{D}_{2}-{B}_{2}{C}_{2}-{B}_{2}{D}_{2}-{C}_{2}{D}_{2}\\
-{A}_{2}{B}_{1}{C}_{1}-{A}_{2}{B}_{1}{D}_{1}-{A}_{2}{C}_{1}{D}_{1}-{A}_{1}{B}_{2}{C}_{1}\\
-{A}_{1}{B}_{2}{D}_{1}-{B}_{2}{C}_{1}{D}_{1}-{A}_{1}{B}_{1}{C}_{2}-{A}_{1}{C}_{2}{D}_{1}\\
-{B}_{1}{C}_{2}{D}_{1}-{A}_{1}{B}_{1}{D}_{2}-{A}_{1}{C}_{1}{D}_{2}-{B}_{1}{C}_{1}{D}_{2}\\
+{A}_{2}{B}_{2}{C}_{2}+{A}_{2}{B}_{2}{D}_{2}+{A}_{2}{C}_{2}{D}_{2}+{B}_{2}{C}_{2}{D}_{2}\\
+{A}_{1}{B}_{1}{C}_{1}{D}_{1}-{A}_{1}{B}_{2}{C}_{1}{D}_{1}-{A}_{2}{B}_{2}{C}_{2}{D}_{2}\\
-{A}_{1}{B}_{1}{C}_{2}{D}_{1}-{A}_{1}{B}_{1}{C}_{1}{D}_{2}-{A}_{2}{B}_{1}{C}_{1}{D}_{1}|/24\leq 1,\label{Hardy4}\\
 \end{split}
\end{equation}
It is violated by the four-qubit GHZ state by a factor of 1.0690, it is violated maximally by a certain state by a factor of 1.1665. On account of noise, we consider a mixed state similar to the three-qubit case as defined below (6). Here $\rho_{\rm{noise}}=\frac{1}{16}\mathbb{I}^{\otimes 4}$ for four qubits, the threshold visibility $V$ is 0.7071 for the GHZ state.
The homogenized inequality can be written as
\begin{equation}
\begin{split}
|20{A}_{0}{B}_{0}{C}_{0}{D}_{0}-2({A}_{1}{B}_{0}{C}_{0}{D}_{0}+{A}_{0}{B}_{1}{C}_{0}{D}_{0}\\
+{A}_{0}{B}_{0}{C}_{1}{D}_{0}+{A}_{0}{B}_{0}{C}_{0}{D}_{1})-{A}_{1}{B}_{1}{C}_{0}{D}_{0}\\
-{A}_{1}{B}_{0}{C}_{1}{D}_{0}-{A}_{1}{B}_{0}{C}_{0}{D}_{1}-{A}_{0}{B}_{1}{C}_{1}{D}_{0}\\
-{A}_{0}{B}_{1}{C}_{0}{D}_{1}-{A}_{0}{B}_{0}{C}_{1}{D}_{1}-{A}_{2}{B}_{1}{C}_{0}{D}_{0}\\
-{A}_{2}{B}_{0}{C}_{1}{D}_{0}-{A}_{2}{B}_{0}{C}_{0}{D}_{1}-{A}_{1}{B}_{2}{C}_{0}{D}_{0}\\
-{A}_{0}{B}_{2}{C}_{1}{D}_{0}-{A}_{0}{B}_{2}{C}_{0}{D}_{1}-{A}_{1}{B}_{0}{C}_{2}{D}_{0}\\
-{A}_{0}{B}_{1}{C}_{2}{D}_{0}-{A}_{0}{B}_{0}{C}_{2}{D}_{1}-{A}_{1}{B}_{0}{C}_{0}{D}_{2}\\
-{A}_{0}{B}_{1}{C}_{0}{D}_{2}-{A}_{0}{B}_{0}{C}_{1}{D}_{2}-{A}_{2}{B}_{2}{C}_{0}{D}_{0}\\
-{A}_{2}{B}_{0}{C}_{2}{D}_{0}-{A}_{2}{B}_{0}{C}_{0}{D}_{2}-{A}_{0}{B}_{2}{C}_{2}{D}_{0}\\
-{A}_{0}{B}_{2}{C}_{0}{D}_{2}-{A}_{0}{B}_{0}{C}_{2}{D}_{2}-{A}_{2}{B}_{1}{C}_{1}{D}_{0}\\
-{A}_{2}{B}_{1}{C}_{0}{D}_{1}-{A}_{2}{B}_{0}{C}_{1}{D}_{1}-{A}_{1}{B}_{2}{C}_{1}{D}_{0}\\
-{A}_{1}{B}_{2}{C}_{0}{D}_{1}-{A}_{0}{B}_{2}{C}_{1}{D}_{1}-{A}_{1}{B}_{1}{C}_{2}{D}_{0}\\
-{A}_{1}{B}_{0}{C}_{2}{D}_{1}-{A}_{0}{B}_{1}{C}_{2}{D}_{1}-{A}_{1}{B}_{1}{C}_{0}{D}_{2}\\
-{A}_{1}{B}_{0}{C}_{1}{D}_{2}-{A}_{0}{B}_{1}{C}_{1}{D}_{2}+{A}_{2}{B}_{2}{C}_{2}{D}_{0}\\
+{A}_{2}{B}_{2}{C}_{0}{D}_{2}+{A}_{2}{B}_{0}{C}_{2}{D}_{2}+{A}_{0}{B}_{2}{C}_{2}{D}_{2}\\
+{A}_{1}{B}_{1}{C}_{1}{D}_{1}-{A}_{1}{B}_{2}{C}_{1}{D}_{1}-{A}_{2}{B}_{2}{C}_{2}{D}_{2}\\
-{A}_{1}{B}_{1}{C}_{2}{D}_{1}-{A}_{1}{B}_{1}{C}_{1}{D}_{2}-{A}_{2}{B}_{1}{C}_{1}{D}_{1}|/24\leq 1,\label{G-Hardy4}
 \end{split}
\end{equation}
\par
It is violated maximally by the four-qubit GHZ state by a factor of 1.9524 (see Fig.~1),  the threshold visibility $V$ is 0.5121.
\par

\section{BELL INEQUALITIES WITH NO QUANTUM VIOLATION}
The quantum correlations (QC) are in general more stronger than classical correlations (CC). Augusiak \emph{et al.}~\cite{UPS1} showed how unextendable product bases (UPBs) that satisfy a given requirement give rise to a family of tight Bell inequalities without quantum violation. In these situations, QC and CC perform equally well in information tasks, while the supraquantum nonsignaling correlations do provide an advantage over CC. Thus such inequalities pinpoint the facet of polytope that separate quantum and supraquantum correlations, and provide better understandings of different sets of correlations which serve as valuable information resources.

Bell inequality with non-negative weights ${q}_{j}$, which always can be assumed to obey $0\leq {q}_{j}\leq 1$.
\begin{eqnarray}
\sum_{j}^{}{q}_{j}p({\emph{\textbf{a}}}_{j}|{\emph{\textbf{x}}}_{j})\leq \max\left\{{q}_{j}\right\},
\end{eqnarray}

From the initial set of orthogonal product vectors, Augusiak \emph{et al.} proved that all these inequalities are not violatd by QC~\cite{UPS1}~\cite{UPS2}.
By local unitaries and permutations of particles, all UPBs can be brought to ${S}_{(i)}^{0}\equiv {S}_{0}=\left\{\left|0\right\rangle,\left|1\right\rangle\right\}$ and ${S}_{(i)}^{1}\equiv {S}_{1}=\left\{\left|e\right\rangle,\left|\overline{e}\right\rangle\right\}(i=1,2,3).$ We assign conditional probabilities in the following way: $\left|000\right\rangle\rightarrow p(000|000), \left|1\overline{e}e\right\rangle\rightarrow p(110|011), \left|e1\overline{e}\right\rangle\rightarrow p(011|101), \left|\overline{e}e1\right\rangle\rightarrow p(101|110).$ Then, we can get the tight Bell inequality with no quantum violation found in~\cite{Sliwa} and~\cite{Guess}.
For odd $n$, the inequality can be written as
\begin{eqnarray}
\sum_{k=0}^{(n-1)/2}\sum_{{i}_{1}<\ldots<{i}_{2k}=1}^{n} {T}_{{i}_{1}\ldots{i}_{2k}}p(\textbf{0}|\textbf{0}) \leq 1,
\end{eqnarray}
and for even $n$
\begin{equation}
\sum_{k=0}^{(n-2)/2}\sum_{{i}_{1}<\ldots<{i}_{2k}=2}^{n}{T}_{{i}_{1}\ldots{i}_{2k}}[p(\textbf{0}|\textbf{0})+p(0\ldots01|10\ldots0)]
\leq 1.
\end{equation}
Here $\textbf{0}=(0,\ldots,0)$, and ${T}_{{i}_{1}\ldots{i}_{2k}}$ denotes a filp $(0\leftrightarrow1)$ of input bits and output bits at positions ${i}_{1},\ldots,{i}_{2k}$ and ${i}_{1}-1,\ldots,{i}_{2k}-1$ (if ${i}_{j}=1$, then ${i}_{j}-1=n$), respectively.
\par
For $n=3$, we obtain the inequality
\begin{equation}
p(000|000)+p(101|110)+p(011|101)+p(110|011)\leq 1.\label{No-QN-3}
\end{equation}
\par
For $n=4$,
\begin{equation}
\begin{split}
p(0000|0000)+p(0001|1000)+p(0110|0011)\\
+p(0111|1011)+p(1010|0101)+p(1011|1101)\\
+p(1100|0110)+p(1101|1110)\leq 1.\label{No-QN-4}
\end{split}
\end{equation}

\begin{figure}
\begin{center}
\includegraphics[width=7.1cm]{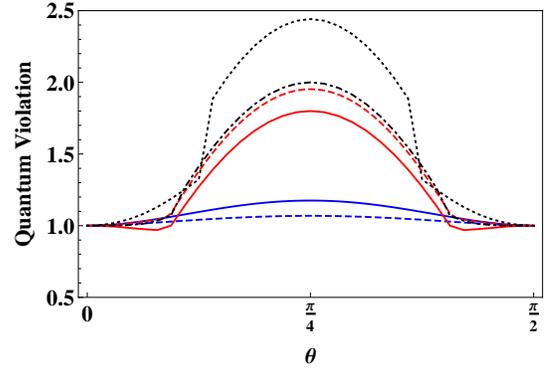}
\end{center}
\caption{The quantum violation for the generalized GHZ state. The blue solid, red solid, blue dash, red dash, black dot-dash, and black dot curves correspond to inequalities (\ref{Hardy3}), (\ref{G-Hardy3}), (\ref{Hardy4}), (\ref{G-Hardy4}), (\ref{QN3}), and (\ref{QN4}), respectively. } \label{fig1}
\end{figure}

\par

Rewriting (\ref{No-QN-3}),
 the symbols in inequalities of probability $0\rightarrow1$ and $1\rightarrow2$ into the correlation function, we obtain
\begin{equation}
\begin{split}
 |{A}_{1}{B}_{1}-{A}_{2}{B}_{1}+{A}_{1}{B}_{2}-{A}_{2}{B}_{2}+{A}_{1}{C}_{1}+{A}_{2}{C}_{1}\\
 +{B}_{1}{C}_{1}-{B}_{2}{C}_{1}-{B}_{2}{C}_{2}-{A}_{1}{C}_{2}-{A}_{2}{C}_{2}+{B}_{1}{C}_{2}\\
 +{A}_{1}{B}_{1}{C}_{1}+{A}_{2}{B}_{2}{C}_{1}+{A}_{2}{B}_{1}{C}_{2}+{A}_{1}{B}_{2}{C}_{2}|/4\leq 1.
 \end{split}
\end{equation}
This inequality cannot be violated in quantum mechanics. However, let us see its homogenized inequality:
\begin{equation}
\begin{split}
  |{A}_{1}{B}_{1}{C}_{0}-{A}_{2}{B}_{1}{C}_{0}+{A}_{1}{B}_{2}{C}_{0}-{A}_{2}{B}_{2}{C}_{0}\\
 +{A}_{1}{B}_{0}{C}_{1}+{A}_{2}{B}_{0}{C}_{1}+{A}_{0}{B}_{1}{C}_{1}-{A}_{0}{B}_{2}{C}_{1}\\
 -{A}_{0}{B}_{2}{C}_{2}-{A}_{1}{B}_{0}{C}_{2}-{A}_{2}{B}_{0}{C}_{2}+{A}_{0}{B}_{1}{C}_{2}\\
 +{A}_{1}{B}_{1}{C}_{1}+{A}_{2}{B}_{2}{C}_{1}+{A}_{2}{B}_{1}{C}_{2}+{A}_{1}{B}_{2}{C}_{2}|/4\leq 1,\label{QN3}
 \end{split}
\end{equation}
which is equivalent to the inequality in Ref.~\cite{Wiesniak}. It is violated maximally by the three-qubit GHZ state by a factor of 2 (see Fig.~1).

For $n=4$
\begin{equation}
\begin{split}
 |2{A}_{1}{C}_{1}+2{B}_{1}{C}_{1}-2{A}_{1}{C}_{2}+2{A}_{2}{C}_{1}-2{A}_{2}{C}_{2}\\
 +2{B}_{1}{C}_{2}-2{B}_{2}{C}_{1}-2{B}_{2}{C}_{2}+{A}_{1}{B}_{1}{C}_{1}+{A}_{2}{B}_{1}{C}_{1}\\
 +{A}_{1}{B}_{2}{C}_{1}+{A}_{2}{B}_{2}{C}_{1}+{A}_{1}{B}_{1}{C}_{2}+{A}_{2}{B}_{1}{C}_{2}+{A}_{1}{B}_{2}{C}_{2}\\
 +{A}_{2}{B}_{2}{C}_{2}+{A}_{1}{B}_{1}{D}_{1}-{A}_{2}{B}_{1}{D}_{1}+{A}_{1}{B}_{2}{D}_{1}-{A}_{2}{B}_{2}{D}_{1}\\
 +{A}_{1}{C}_{1}{D}_{1}-{A}_{2}{C}_{1}{D}_{1}-{A}_{1}{C}_{2}{D}_{1}+{A}_{2}{C}_{2}{D}_{1}-{A}_{1}{B}_{1}{D}_{2}\\
 +{A}_{2}{B}_{1}{D}_{2}-{A}_{1}{B}_{2}{D}_{2}+{A}_{2}{B}_{2}{D}_{2}+{A}_{1}{C}_{1}{D}_{2}-{A}_{2}{C}_{1}{D}_{2}\\
 -{A}_{1}{C}_{2}{D}_{2}+{A}_{2}{C}_{2}{D}_{2}+{A}_{1}{B}_{1}{C}_{1}{D}_{1}-{A}_{2}{B}_{1}{C}_{1}{D}_{1}\\
 +{A}_{1}{B}_{2}{C}_{2}{D}_{1}-{A}_{2}{B}_{2}{C}_{2}{D}_{1}+{A}_{1}{B}_{2}{C}_{1}{D}_{2}-{A}_{2}{B}_{2}{C}_{1}{D}_{2}\\
 +{A}_{1}{B}_{1}{C}_{2}{D}_{2}-{A}_{2}{B}_{1}{C}_{2}{D}_{2}|/8\leq 1,
 \end{split}
\end{equation}
whose homogenized inequality can be written as
\begin{equation}
\begin{split}
 |2{A}_{1}{B}_{0}{C}_{1}{D}_{0}+2{A}_{2}{B}_{0}{C}_{1}{D}_{0}+2{A}_{0}{B}_{1}{C}_{1}{D}_{0}\\
 -2{A}_{0}{B}_{2}{C}_{1}{D}_{0}-2{A}_{1}{B}_{0}{C}_{2}{D}_{0}-2{A}_{2}{B}_{0}{C}_{2}{D}_{0}\\
 +2{A}_{0}{B}_{1}{C}_{2}{D}_{0}-2{A}_{0}{B}_{2}{C}_{2}{D}_{0}+{A}_{1}{B}_{1}{C}_{1}{D}_{0}\\
 +{A}_{2}{B}_{1}{C}_{1}{D}_{0}+{A}_{1}{B}_{2}{C}_{1}{D}_{0}+{A}_{2}{B}_{2}{C}_{1}{D}_{0}\\
 +{A}_{1}{B}_{1}{C}_{2}{D}_{0}+{A}_{2}{B}_{1}{C}_{2}{D}_{0}+{A}_{1}{B}_{2}{C}_{2}{D}_{0}\\
 +{A}_{2}{B}_{2}{C}_{2}{D}_{0}+{A}_{1}{B}_{1}{C}_{0}{D}_{1}-{A}_{2}{B}_{1}{C}_{0}{D}_{1}\\
 +{A}_{1}{B}_{2}{C}_{0}{D}_{1}-{A}_{2}{B}_{2}{C}_{0}{D}_{1}+{A}_{1}{B}_{0}{C}_{1}{D}_{1}\\
 -{A}_{2}{B}_{0}{C}_{1}{D}_{1}-{A}_{1}{B}_{0}{C}_{2}{D}_{1}+{A}_{2}{B}_{0}{C}_{2}{D}_{1}\\
 -{A}_{1}{B}_{1}{C}_{0}{D}_{2}+{A}_{2}{B}_{1}{C}_{0}{D}_{2}-{A}_{1}{B}_{2}{C}_{0}{D}_{2}\\
 +{A}_{2}{B}_{2}{C}_{0}{D}_{2}+{A}_{1}{B}_{0}{C}_{1}{D}_{2}-{A}_{2}{B}_{0}{C}_{1}{D}_{2}\\
 -{A}_{1}{B}_{0}{C}_{2}{D}_{2}+{A}_{2}{B}_{0}{C}_{2}{D}_{2}+{A}_{1}{B}_{1}{C}_{1}{D}_{1}\\
 -{A}_{2}{B}_{1}{C}_{1}{D}_{1}+{A}_{1}{B}_{2}{C}_{2}{D}_{1}-{A}_{2}{B}_{2}{C}_{2}{D}_{1}\\
 +{A}_{1}{B}_{2}{C}_{1}{D}_{2}-{A}_{2}{B}_{2}{C}_{1}{D}_{2}+{A}_{1}{B}_{1}{C}_{2}{D}_{2}\\
 -{A}_{2}{B}_{1}{C}_{2}{D}_{2}|/8\leq 1.\label{QN4}
 \end{split}
\end{equation}
It is violated maximally by the four-qubit GHZ state by a factor of 2.4413 (see Fig.~1).
\par

\begin{figure}
\begin{center}
\includegraphics[width=7.1cm]{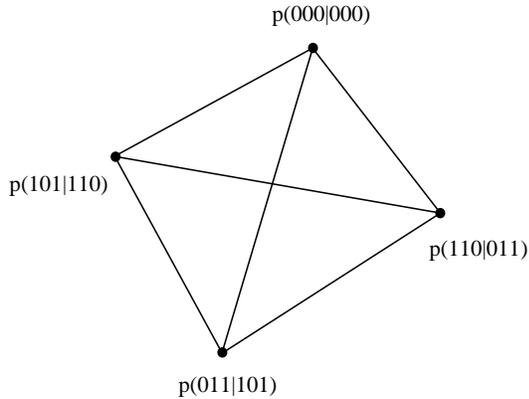}
\end{center}
\caption{The exclusivity graph for (\ref{No-QN-3}). The sum of probabilities of pairwise exclusive events cannot exceed 1.} \label{fig2}
\end{figure}

%\begin{figure}
%\begin{center}
%%\includegraphics[width=7.1cm]{fig3.eps}
%\end{center}
%\caption{The exclusivity graph for (\ref{QN3}). This is not a complete graph.} \label{fig2}
%\end{figure}

\section{DISCUSSION AND CONCLUSIONS}
\par
To summarize, we have studied the influence of homogenization on enhancing the quantum nonlocality of two families of Bell inequalities. Through the homogenization, the quantum violation of the Hardy inequality has become stronger for the GHZ state, while the parameter domain for violation regime of the generalized GHZ state has been narrowed. On the other hand, quantum violation has appeared again for the family of tight Bell inequalities without quantum violation. The reason for this is that a tight Bell inequality without quantum violation can be represented by a \emph{complete} graph~\cite{cabello2014} (see Fig.~\ref{fig2}), i.e., event probabilities are pairwise exclusive, so that, according to graph theory, the independence number is equal to the Lov\'asz number~\cite{Lovasz}, indicating the coincidence of the classical upperbound with the quantum maximum. However, by homogenization, it is transformed into a full-correlation form, whose corresponding graph is no longer complete. In this regard, the independence number is in general less than than the Lov\'asz number, rendering a quantum violation possible.

\begin{acknowledgments}
This work is supported by the National Basic Research Program
(973 Program) of China (Grant Nos.  2012CB921900, 2011CBA00200, and
2011CB921200) and the NSF of China (Grant Nos. 11175089, 11475089, 10974193,
11275182, and 60921091). J.L.C. is partly supported by the
National Research Foundation and the Ministry of
Education, Singapore.
\end{acknowledgments}

\end{document}